\documentclass[twocolumn,trackchanges]{aastex62}

\usepackage{graphicx}

\begin{document}

\title{Methanol and its relation to the water snowline in the disk around the young outbursting star V883 Ori}

\correspondingauthor{Merel L.R. van 't Hoff}
\email{vthoff@strw.leidenuniv.nl}

\author{Merel L.R. van 't Hoff}
\affil{Leiden Observatory, Leiden University, P.O. box 9513, 2300 RA Leiden, The Netherlands}

\author{John J. Tobin}
\affiliation{Leiden Observatory, Leiden University, P.O. box 9513, 2300 RA Leiden, The Netherlands}
\affiliation{Homer L. Dodge Department of Physics and Astronomy, University of Oklahoma, 440 W. Brooks Street, Norman, OK 73019, USA}

\author{Leon Trapman}
\affiliation{Leiden Observatory, Leiden University, P.O. box 9513, 2300 RA Leiden, The Netherlands}

\author{Daniel Harsono}
\affiliation{Leiden Observatory, Leiden University, P.O. box 9513, 2300 RA Leiden, The Netherlands}

\author{Patrick D. Sheehan}
\affiliation{Homer L. Dodge Department of Physics and Astronomy, University of Oklahoma, 440 W. Brooks Street, Norman, OK 73019, USA}

\author{William J. Fischer}
\affiliation{Space Telescope Science Institute, 3700 San Martin Drive, Baltimore, MD 21218, USA}

\author{S. Thomas Megeath}
\affiliation{Ritter Astrophysical Research Center, Department of Physics and Astronomy, University of Toledo, 2801 West Bancroft Street, Toledo, OH 43606, USA}

\author{Ewine F. van Dishoeck}
\affiliation{Leiden Observatory, Leiden University, P.O. box 9513, 2300 RA Leiden, The Netherlands}
\affiliation{Max-Planck-Institut f\"{u}r Extraterrestrische Physik, Giessenbachstrasse 1, 85748 Garching, Germany}



\begin{abstract}

We report the detection of methanol in the disk around the young outbursting star V883 Ori with the Atacama Large Millimeter/submillimeter Array (ALMA). Four transitions are observed with upper level energies ranging between 115 and 459 K. The emission is spatially resolved with the 0.14$\arcsec$ beam and follows the Keplerian rotation previously observed for C$^{18}$O. Using a rotational diagram analysis, we find a disk-averaged column density of $\sim10^{17}$ cm$^{-2}$ and a rotational temperature of \mbox{$\sim90-100$ K}, suggesting that the methanol has thermally desorbed from the dust grains. We derive outer radii between 120 and 140 AU for the different transitions, compared to the 360 AU outer radius for C$^{18}$O. Depending on the exact physical structure of the disk, the methanol emission could originate in the surface layers beyond the water snowline. Alternatively, the bulk of the methanol emission originates inside the water snowline, which can then be as far out as $\sim$100 AU, instead of 42 AU as was previously inferred from the continuum opacity. 
In addition, these results show that outbursting young stars like V883 Ori are good sources to study the ice composition of planet forming material through thermally desorbed complex molecules, which have proven to be hard to observe in more evolved protoplanetary disks. 

\end{abstract}



\section{Introduction} \label{sec:introduction}

Snowlines in disks around young stars mark the midplane locations where molecular species freeze out from the gas phase onto dust grains. The most important snowline for planet formation is the water snowline. Planetesimal formation is expected to be significantly enhanced at this snowline because the bulk of the ice mass is in water ice \citep[e.g.,][]{Stevenson1988,Ros2013,Schoonenberg2017a}. In addition, the elemental composition of both the gas and ice changes across this snowline since water is a major carrier of oxygen. The bulk composition of planets therefore depends on their formation location with respect to the water snowline \citep[e.g.,][]{Oberg2011,Madhusudhan2014,Eistrup2018}.

\begin{table*}[ht!]
\caption{Overview of the molecular line observations toward V883 Ori.
\label{tab:Lineparameters}} 
\centering
\begin{tabular}{l c c c c c c c c c}
    \hline\hline
    Species & Transition & Frequency & $A_{\rm{ul}}$ & $E_{\rm{up}}/k$ & $g_{\rm{up}}$ & $F_{\rm{peak}}$ & $F_{\rm{int}}$\tablenotemark{a} & $R_{\rm{out}}$ \\ 
    & & (GHz) & (s$^{-1}$) & (K) & & (mJy beam$^{-1}$) & (Jy km s$^{-1}$) & (AU) \\
    \hline 
    C$^{18}$O		& $2-1$			& 219.560354 & 6.03$\times$10$^{-7}$ & 16  & 3  & 117 $\pm$ 6 & 1.1 $\pm$ 0.08 & 361 $\pm$ 23  \\
    CH$_3$OH	(A$^-$)	& $5_4-6_3$		& 346.202719 & 2.12$\times$10$^{-5}$ & 115 & 11 & \hspace{0.3cm}105 $\pm$ 21\tablenotemark{b} & \hspace{0.15cm}1.6 $\pm$ 0.18\tablenotemark{b} & \hspace{0.15cm}142 $\pm$ 27\tablenotemark{b} \\
    CH$_3$OH	(A$^+$)	& $5_4-6_3$		& 346.204271 & 2.12$\times$10$^{-5}$ & 115 & 11 & \hspace{0.3cm}105 $\pm$ 21\tablenotemark{b} & \hspace{0.15cm}1.6 $\pm$ 0.18\tablenotemark{b} & \hspace{0.15cm}142 $\pm$ 27\tablenotemark{b}  \\
    CH$_3$OH	(A$^-$)	& $16_1-15_2$	& 345.903916 & 8.78$\times$10$^{-5}$ & 333 & 33 & \hspace{0.15cm}108 $\pm$ 21 & 1.1 $\pm$ 0.18 & 136 $\pm$ 18  \\
    CH$_3$OH	(E2)		& $18_3-17_4$	& 345.919260 & 7.10$\times$10$^{-5}$ & 459 & 37 & \hspace{0.3cm}77 $\pm$ 21 & 0.5 $\pm$ 0.18 & 117 $\pm$ 26\\
    \hline
\end{tabular}
\tablenotetext{a}{Within a 1.0$\arcsec$ aperture for C$^{18}$O and within a 0.6$\arcsec$ aperture for CH$_3$OH.} 
\tablenotetext{b}{Between 0.5 and 7.0 km s$^{-1}$, that is, for both lines combined.}
\end{table*}
%

For water, the transition from ice to gas occurs when the temperature exceeds roughly 100 K \citep{Fraser2001}. This places the snowline at a few AU from the star in protoplanetary disks, making it hard to observe. However, heating is temporarily enhanced during protostellar accretion bursts, which causes ices to sublimate out to larger radii. After such a burst, the circumstellar dust cools rapidly \citep{Johnstone2013}, while for the molecules it takes much longer to freeze back onto the grains \citep{Rodgers2003}. As a result snowlines are shifted away from the star \citep{Lee2007,Visser-Bergin2012,Vorobyov2013,Visser2015}. This has been observed for CO toward a sample of protostars \citep{Jorgensen2015,Frimann2017}. 

V883 Ori is a FU Orionis object in the Orion A L1641 molecular cloud ($d \sim 400$ pc; \citealt{Kounkel2017}) with a bolometric luminosity of $\sim218 L_{\sun}$ \citep{Strom1993,Furlan2016}. Although the onset of the V883 Ori outburst was not directly observed, evidence for an ongoing outburst that began before 1888 comes from its associated reflection nebula \citep{Pickering1890} and the similarity of its near-IR spectrum to that of FU Ori \citep{Connelley2018}. The 1.3 $M_{\sun}$ star is surrounded by a $\gtrsim$0.3 M$_{\sun}$ rotationally supported disk \citep{Cieza2016,Cieza2018} still embedded in its envelope. 

The location of the water snowline in V883 Ori was inferred from a change in the continuum opacity at 42 AU \citep{Cieza2016}, which may be due to a pileup of dust interior to the snowline \citep{Birnstiel2010,Banzatti2015,Pinilla2016}, or water evaporation and re-coagulation of silicates \citep{Schoonenberg2017b}. However, the origin of various structures seen in continuum emission of disks is still heavily debated and radial discontinuities in the spectral index are not necessarily related to snowlines \citep{vanTerwisga2018}. Molecular observations are thus needed to confirm or refute the snowline location. Unfortunately, water is hard to observe from the ground and warm water ($T \gtrsim 100$ K) has not yet been observed in young disks (Harsono et al., in prep.), so observing the snowline directly is difficult. A complementary approach is to observe other molecules whose distribution can be related to the snowline. Methanol (CH$_3$OH) is generally used to probe the $\gtrsim$ 100 K region in hot cores \citep{herbst2009}, because its volatility is similar to that of water \citep[e.g.,][]{Brown2007}. 

We serendipitously detected spatially resolved methanol emission in the V883 Ori disk with the VLA/ALMA Nascent Disk And Multiplicity (VANDAM) Orion survey that aims to characterize the embedded disks in Orion (PI: Tobin, Tobin et al., in prep.). Analysis of the methanol observations and comparison with earlier C$^{18}$O observations shows that the methanol is thermally desorbed and suggests that the water snowline can be as far out as $\sim100$ AU. 

\vspace{0.5cm}

\section{Observations} \label{sec:observations}

V883 Ori was observed on 2016 September 6 and 2017 July 19 in Band 7 as part of ALMA Cycle 4 project 2015.1.00041.S. The total on-source integration time was 54 seconds, and baselines between 16.7 m and 3697 m were covered. The correlator setup consisted of two low spectral resolution (31.25~MHz) 1.875~GHz continuum windows centered at 333~GHz and 344~GHz, a 234.375 MHz spectral window (122 kHz resolution) centered at 330.6~GHz, and a 937.5~MHz spectral window (488 kHz $\approx$ 0.42 km s$^{-1}$ resolution) centered at 345.8~GHz. The latter spectral window contained the methanol transitions. 

The bandpass calibrator was J0510+1800 for the 2016 observations and J0522-3627 for the 2017 observations. The absolute flux calibrators were J0510+1800, J0522-3627 and J0750+1231 for the respective executions, and J0541-0541 was used as complex gain calibrator for all executions. The data were reduced manually by the Dutch Allegro ARC Node to properly account for variation of quasar J0510+1800. Following the standard calibration, phase-only self-calibration was performed on the continuum data using version 4.7.2 of the Common Astronomy Software Application (CASA, \citealt{McMullin2007}). The self-calibration solutions were also applied to the spectral line data. The line data was imaged after continuum subtraction, using the CASA task \textit{clean} with natural weighting and a velocity resolution of 0.5 km s$^{-1}$. This resulted in a synthesized beam size of $0.13\arcsec \times 0.14\arcsec$ and a rms of 21 mJy beam$^{-1}$ per channel. 

\begin{figure*}[ht!]
\centering
\includegraphics[width=\textwidth,trim={0cm 11.5cm 0cm 1.7cm},clip]{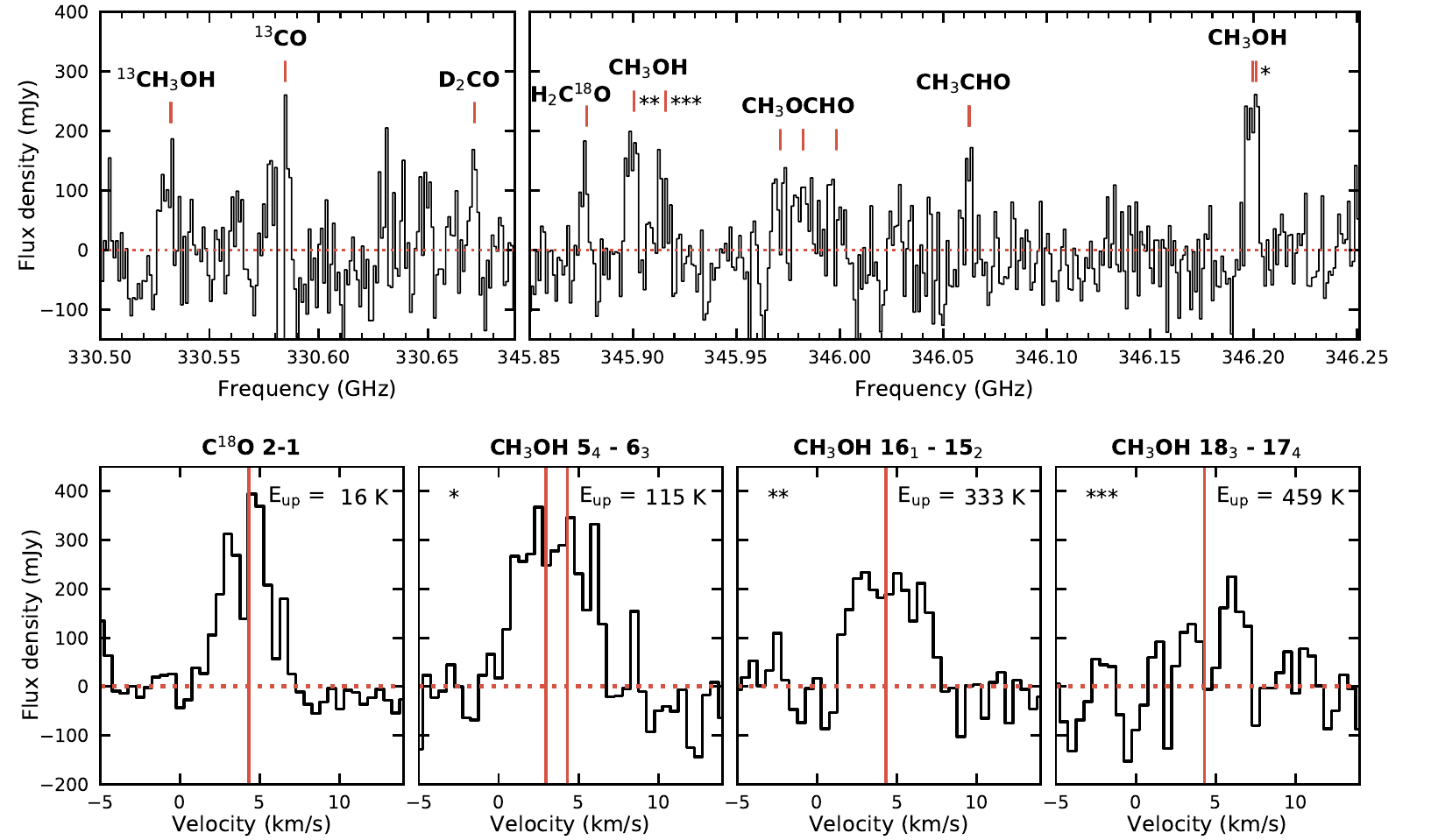}
\caption{Spectra extracted within a 0.6$\arcsec$ circular aperture (1.0$\arcsec$ for C$^{18}$O) toward V883 Ori.  The \textit{top panels} display part of the spectral windows centered at 330.6 GHz and 345.8 GHz binned to a resolution of 1.10 and 1.16 MHz, respectively (1.0 km~s$^{-1}$). In addition, an outer taper of 2000 k$\lambda$ was applied before imaging. The transitions for CH$_3$OH and a few other key species are labeled and marked with a vertical red line. The rms from a line-free region is $\sim$50 mJy. The \textit{bottom panels} show the C$^{18}$O and CH$_3$OH spectra at 0.5 km s$^{-1}$ resolution. The asterisks in the top left corner of the CH$_3$OH panels correspond to the asterisks in the \textit{top right panel}. The vertical red lines indicate the line centers \mbox{($v_{\rm{sys}}$ = 4.3 km s$^{-1}$)}.}
\label{fig:Spectra}
\end{figure*}


V883 Ori was also observed in Band 6 (project 2013.1.00710.S). In these observations the correlator was configured to have one baseband centered on the C$^{18}$O $J=2-1$ transition at 219.560 GHz. The C$^{18}$O data reduction is described by \citet{Cieza2016}. The resulting image has a synthesized beam of $0.23\arcsec \times 0.30\arcsec$ and a rms of 9 mJy beam$^{-1}$ in 0.5 km s$^{-1}$ channels.

\vspace{1cm}

\section{Results} \label{sec:results}

\subsection{Detection of warm methanol in the disk}

The spectral setting of the ALMA Band 7 observations covers four CH$_3$OH lines with upper level energies ranging from 115 to 459 K (see Table~\ref{tab:Lineparameters}). Spectra centered at the corresponding rest frequencies are presented in Figure~\ref{fig:Spectra}. All four lines are detected between $\sim$0.5 and $\sim$7.0 km s$^{-1}$, but the two nearby $5_4-6_3$ transitions from A$^+$ and A$^-$ CH$_3$OH are blended. The peak signal-to-noise in the 0.5 km s$^{-1}$ channels is $\sim$4 for the highest energy transition ($18_3-17_4$) and $\sim$5 for the other lines. 

In addition to the CH$_3$OH transitions, several other lines are marginally detected (Figure~\ref{fig:Spectra}, top panels). Using the Jet Propulsion Laboratory (JPL, \citealt{Pickett1998}) and the Cologne Database for Molecular Spectroscopy (CDMS, \citealt{Muller2001}) databases through Splatalogue\footnote{www.splatalogue.net} we can assign transitions of $^{13}$CH$_3$OH, the formaldehyde isotopologues H$_2$C$^{18}$O and D$_2$CO, methyl formate (CH$_3$OCHO) and acetaldehyde (CH$_3$CHO). The peak signal-to-noise of these transitions are $\sim4-5 \sigma$ in 1.0 km s$^{-1}$ channels, but unambigious identification requires detection of more lines. 

\begin{figure*}[ht!]
\centering
\includegraphics[width=\textwidth,trim={0cm 12cm 0cm 1cm},clip]{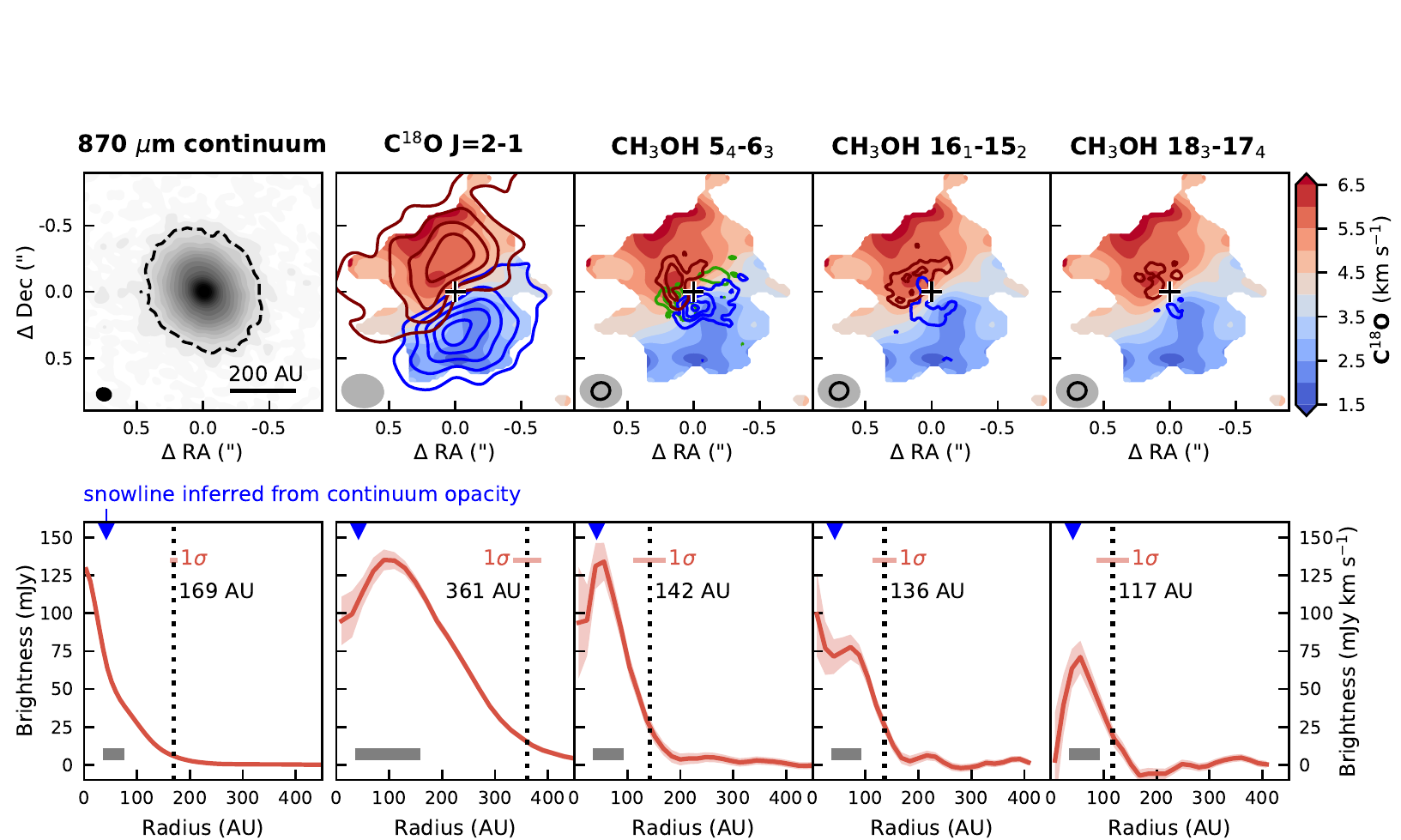}
\caption{Continuum image toward V883 Ori at 870 $\mu$m (\textit{top left panel}), and integrated intensity (moment zero) maps for C$^{18}$O and CH$_3$OH (solid contours) overlaid on the C$^{18}$O intensity-weighted mean velocity (moment one) map (color scale, \textit{top right panels}). The dashed black contour marks the 5$\sigma$ (2.6 mJy beam$^{-1}$) level for the continuum. The red (blue) solid contours show the emission integrated over redshifted (blueshifted) velocity channels containing $> 3\sigma$ emission. The green contours for the CH$_3$OH $5_4 - 6_3$ transitions show emission integrated over velocities between the line centers of the A$^-$ and A$^+$ transitions. For C$^{18}$O, contours are in steps of 3$\sigma$ starting at 3$\sigma$ (11 mJy beam$^{-1}$ km s$^{-1}$). For CH$_3$OH, a Keplerian mask is used, so each moment zero pixel represents the sum of a different number of channels. A signal-to-noise map is therefore calculated using the channel rms and the number of channels used per pixel. Contours are shown in steps of 2$\sigma$ starting at 3$\sigma$. The continuum peak position (RA = 05$^{\rm{h}}$38$^{\rm{m}}$18$\fs$10; Dec = -7$\degr$02$\arcmin$25$\farcs$96) is marked with a cross. The synthesized beams are shown in the lower left corner of the panels, with the continuum beam in black, the C$^{18}$O beam in gray and the CH$_3$OH beams in black contours. The \textit{bottom panels} show the corresponding deprojected and azimuthally averaged radial profiles. Shaded regions denote the 1$\sigma$ uncertainty levels. The dotted line indicates the emission outer radius, that is, the radius containing 90\% of the total integrated flux. The corresponding 1$\sigma$ uncertainty is calculated from the 1$\sigma$ uncertainty on the 90\% flux measurement (horizontal red line). The gray bars (lower left corners) represent the major axis of the restoring beam and the blue triangles (top left corners) mark the snowline location (42 AU) inferred by \citet{Cieza2016} from the continuum opacity.}
\label{fig:MomentsRadialProfiles}
\end{figure*}

The CH$_3$OH channel maps (not shown) display the the same butterfly pattern as C$^{18}$O, typical for a Keplerian rotating disk \citep{Cieza2016}. The kinematics can be more clearly visualized using moment maps. Figure~\ref{fig:MomentsRadialProfiles} (top panels) shows the moment zero (integrated intensity) maps for blue- and redshifted emission of the different CH$_3$OH transitions overlaid on the 1st moment (intensity-weighted mean velocity) map of C$^{18}$O. These moment maps are constructed using a Keplerian mask to enhance the signal-to-noise, that is, only those pixels that are expected to emit for a Keplerian rotating disk are included \citep[see e.g.,][]{Salinas2017,Loomis2018}. The C$^{18}$O 1st moment map shows a clear velocity gradient as expected for a Keplerian rotating disk. The CH$_3$OH emission is more compact than the C$^{18}$O emission, but for all CH$_3$OH transitions a similar velocity gradient is observed with the blueshifted emission spatially coinciding with the blueshifted part of the C$^{18}$O disk and the redshifted emission with the redshifted part. The CH$_3$OH emission thus originates in the disk and not in the surrounding envelope or outflow.

\subsection{Column density and excitation temperature} \label{subsec:rd}

\begin{figure}[t!]
\centering
\includegraphics[width=\textwidth,trim={0.3cm 15.3cm 0cm 2cm},clip]{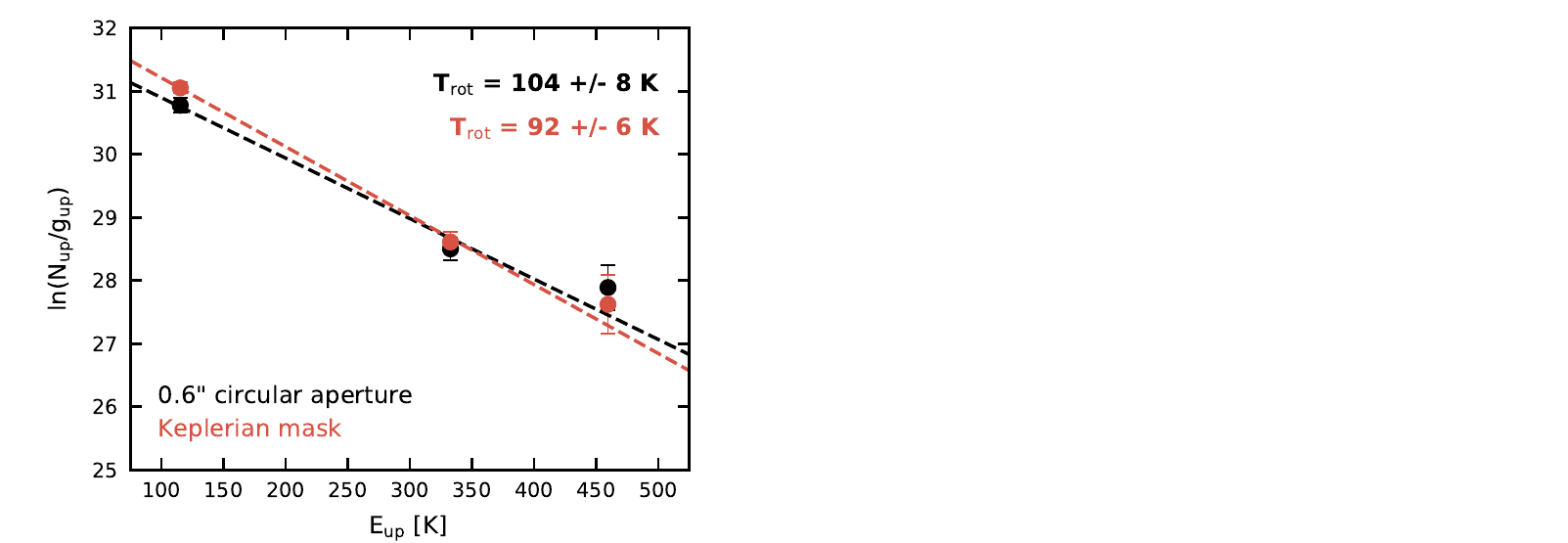}
\caption{CH$_3$OH rotational diagram, constructed from the total integrated flux within a 0.6$\arcsec$ circular aperture (black) or within a Keplerian mask (red). The dashed lines represent the best fit to the data points and the corresponding rotational temperatures are presented in the top right corner.}
\label{fig:RotationDiagram}
\end{figure}

A rotational diagram for CH$_3$OH is presented in Figure ~\ref{fig:RotationDiagram}. The total flux for the two $5_4-6_3$ transitions is divided by two, because the two lines have the same upper level energy and Einstein A coefficient. Fitting a linear function results in a rotation temperature of $104 \pm 8$ K when the flux is extracted within a 0.6$\arcsec$ aperture, and $T_{\rm{rot}} = 92 \pm 6$ K for the Keplerian mask with an outer radius of 225 AU (0.54$\arcsec$). The resulting disk-averaged column densities are $8.9 \pm 1.6 \times 10^{16}$ cm$^{-2}$ and $1.4 \pm 0.2 \times 10^{17}$ cm$^{-2}$, respectively.

If the emission is optically thin and in LTE, the rotational temperature equals the kinetic temperature of the gas. For densities higher than $\sim10^8-10^9$ cm$^{-3}$ the CH$_3$OH excitation temperature is similar to the kinetic temperature \citep[see e.g.,][]{Johnstone2003,Jorgensen2016}, so LTE is a valid assumption in the bulk of the disk where the density is of order $10^9-10^{13}$ cm$^{-3}$. The observed ratios of the integrated fluxes are consistent with optically thin emission based on a LTE calculation, although optically thick emission cannot be completely ruled out with the signal-to-noise of the observations. However, upper levels with an energy of 459 K are hardly populated at temperatures $\lesssim$30 K, and temperatures $\gtrsim$75 K are required for the $18_3-17_4$ flux ($E_{\rm{up}}$ = 459 K) to be more than 30\% of the $16_1-15_2$ flux ($E_{\rm{up}}$ = 333 K). These results thus suggest that we observe methanol that has desorbed thermally.  

A first estimate of the CH$_3$OH abundance can be made by comparing the CH$_3$OH and C$^{18}$O column densities. Using the total flux within a 0.6$\arcsec$ aperture, the C$^{18}$O column density is $\sim 4\times10^{16}$ cm$^{-2}$. Assuming a standard interstellar medium (ISM) $^{16}$O/$^{18}$O ratio of 560 \citep{Wilson1994} and a CO abundance of 10$^{-4}$ with respect to H$_2$, we then derive a CH$_3$OH abundance of $\sim4 \times 10^{-7}$ with respect to H$_2$. This is consistent with observed CH$_3$OH ice abundances of $\sim10^{-6}-10^{-5}$ \citep{Boogert2015}, but not with the gas-phase abundance of $\sim10^{-12}-10^{-11}$ required to explain the TW Hya observations of non-thermally desorbed CH$_3$OH \citep{Walsh2016}.

\subsection{Methanol outer radius} \label{subsec:Rout}

The radial intensity profiles for C$^{18}$O and CH$_3$OH are calculated from the moment zero maps and shown in Figure~\ref{fig:MomentsRadialProfiles}. An inclination of $38^{\circ}$ and a position angle of $32^{\circ}$ east of north \citep{Cieza2016} are used for deprojection and azimuthally averaging. All transitions, except the $16_1-15_2$ CH$_3$OH transition, peak off source. This is likely due to the dust being optically thick in the inner $\sim$40~AU \citep{Cieza2016}. The C$^{18}$O peak is shifted further outward compared to the CH$_3$OH peaks due to the larger beam for C$^{18}$O and possibly optical depth effects. 

We derive outer radii for the emission using a curve-of-growth method \citep[e.g.,][]{Tripathi2017,Ansdell2018}, in which the flux is measured within successively larger photometric apertures until the measured flux is 90\% of the total flux. The same elliptical apertures are used as for the radial intensity profiles. The resulting outer radii ($R_{\rm{out}}$) are listed in Table~\ref{tab:Lineparameters} and indicated in Figure~\ref{fig:MomentsRadialProfiles} (bottom panels). The CH$_3$OH outer radii range between $\sim$117 and $\sim$142 AU for the different transitions, with the outer radius decreasing with upper level energy. The CH$_3$OH outer radii are $\sim 2.5-3.0\times$ smaller than the C$^{18}$O outer radius of $\sim$361 AU. The largest angular scale (LAS) is $\sim4\arcsec$ ($\sim1650$ AU) for the C$^{18}$O observations and $\sim1.5\arcsec$ ($\sim600$ AU) for the CH$_3$OH observations.





\section{Discussion} \label{sec:discussion}

\subsection{Location of the water snowline}

\begin{figure*}[ht!]
\centering
\includegraphics[width=\textwidth,trim={0cm 18.6cm 0cm 1.5cm},clip]{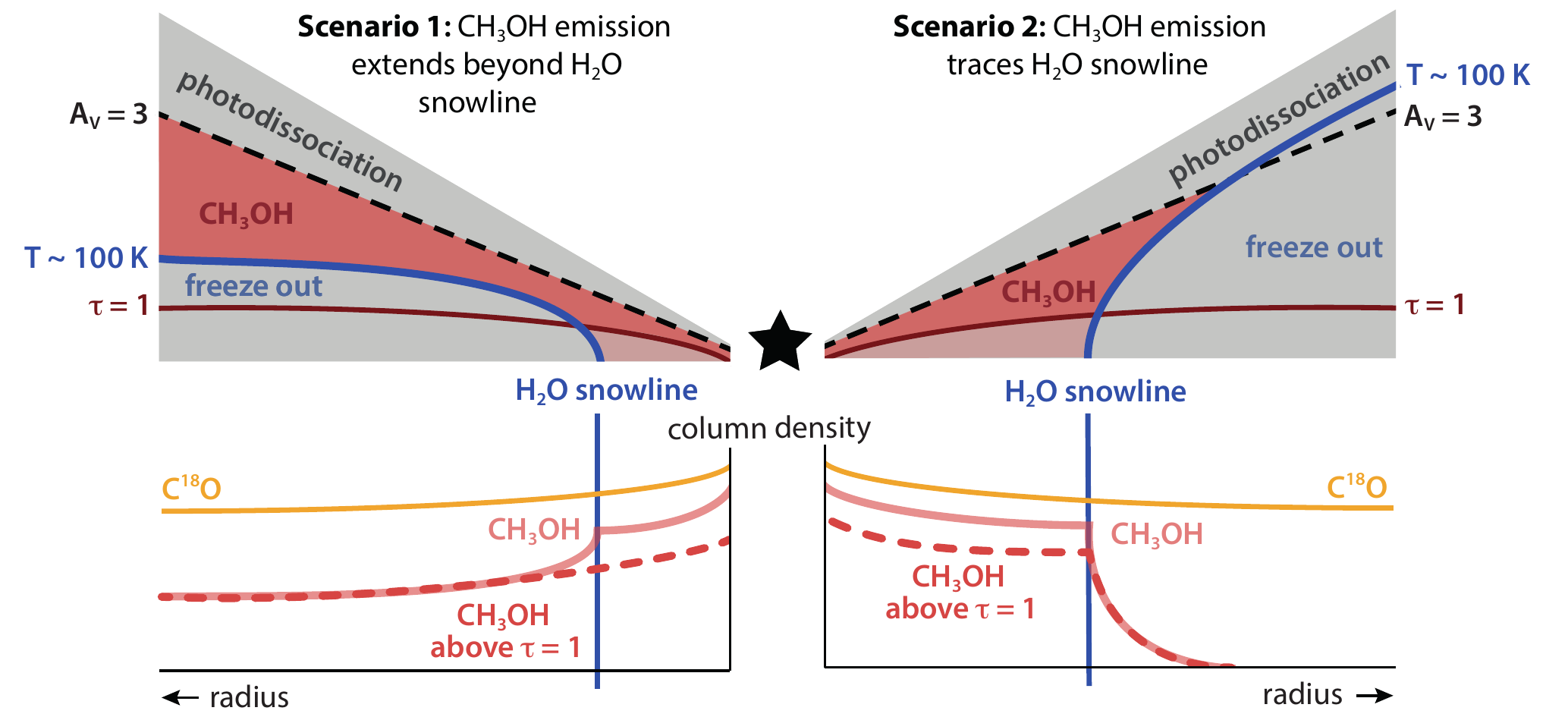}
\caption{Illustration showing two possible scenarios for the distribution of CH$_3$OH with respect to the water snowline (\textit{top}), and the corresponding column density profiles (\textit{bottom}). Distinguishing between the scenarios \textbf{requires} a detailed physical model of the V883 Ori disk. The region where CH$_3$OH is present in the gas phase is highlighted in red. If the CH$_3$OH emission is optically thick, only molecules above the $\tau=1$ surface (red line) are observed. The location of the $\tau=1$ surface depends on both the disk structure and the CH$_3$OH transition. The corresponding column density traced in the optically thick case is shown with a dashed line, while the solid line represents the total CH$_3$OH column. The blue curve indicates the water snow surface at $\sim$100 K, and the vertical blue line the corresponding midplane water snowline. CH$_3$OH desorbs at a similar temperature as water. The dashed black line marks the $A_V=3$ contour above which CH$_3$OH is photodissociated. The C$^{18}$O column density profile is shown in orange.}
\label{fig:Cartoon}
\end{figure*}

The distribution of CH$_3$OH, and hence the relationship between its emission and the water snowline, depends on the physical structure of the disk, as illustrated in Figure~\ref{fig:Cartoon}. CH$_3$OH is present in the gas phase where the temperature exceeds the thermal desorption temperature of $\sim$100 K, and where there is a sufficiently large column of material to shield the UV radiation and prevent photodissociation ($A_V \geq 3$). In the surface layers, the photodissociation timescale is tens of years \citep{Heays2017}, comparable to the outburst duration (10-100 years). This means that the radial extent of the methanol layer higher up in the disk beyond the midplane water snowline is set by the intercept of the snow surface and the $A_V=3$ contour. In addition, the magnitude of the CH$_3$OH column density drop across the water snowline depends on the height of the snow surface; the higher up in the disk the larger the drop. Whether the emission then traces this column density profile depends on the optical depth of both the CH$_3$OH as well as the dust. 

Due to this interplay of several parameters, a detailed physical model of the disk is required to derive the water snowline location from methanol emission. It may thus be possible to see CH$_3$OH emission out to $\sim120-140$AU while the snowline is around 40 AU. This would require for example a water snow surface close to the midplane, and/or optically thick methanol emission (Figure~\ref{fig:Cartoon}, left panel). The vertical temperature structure is strongly dependent on the dust distribution and settling of the large grains can result in a steep vertical temperature profile with the snow surface closer to the midplane than in the case of less grain settling \citep[e.g.,][]{Facchini2017}. However, especially if the emission is optically thin, the bulk of the methanol emission is more likely to originate inside the water snowline (Figure~\ref{fig:Cartoon}, right panel), since the CH$_3$OH column density can drop $\sim$3 orders of magnitude crossing the snowline assuming a constant abundance for the gas-phase CH$_3$OH \citep[see e.g., the simple model for the CO snowline in][]{Qi2013}. Assuming a step function for the column density, this would mean that the snowline in V883 Ori can be as far out as $\sim100-125$ AU, taking into account the 40 AU beam by deconvolving the radial profiles. 

Non-thermal desorption processes are not expected to influence the relationship between CH$_3$OH and the water snowline. Such processes have been invoked to explain the CH$_3$OH emission in TW Hya \citep{Walsh2016}, but this required gas-phase CH$_3$OH outside the CO snowline ($T\lesssim$ 20~K) at an abundance of $\sim10^{-12}-10^{-11}$, several orders of magnitude lower than observed here and expected from ice abundances (Sect.~\ref{subsec:rd}).

Besides detailed modeling, observations of other molecular tracers could put better constraints on the water snowline location. H$^{13}$CO$^+$ has shown to be a promising tracer in the envelope around NGC1333 IRAS2A \citep{vantHoff2018}, because the main destroyer of HCO$^+$ is gas-phase water. 

\vspace{1.0cm}

\subsection{Ice composition of planet forming material}

One of the key questions in planet formation is whether planetary systems inherit their chemical composition from the natal cloud or whether the material is significantly processed en route to the disk. Observations of many complex molecules, including methanol, around young protostars at solar system scales \citep[e.g.,][]{Jorgensen2016} and in comets \citep[e.g.,][]{Mumma2011,LeRoy2015} show that a large complexity is present during both the early as well as the final stages of planet formation. The chemical complexity in protoplanetary disks, however, is hard to probe. Due to the low temperatures ($<$100 K), complex molecules are frozen out onto dust grains at radii larger than a few AU, and ices can only be observed through infrared absorption in edge-on systems. Although alternative desorption processes may get these molecules into the gas phase, as has been shown for water \citep{Hogerheijde2011}, so far only CH$_3$OH and CH$_3$CN have been observed in disks \citep{Oberg2015b,Walsh2016,Bergner2018,Loomis2018}. Moreover, as it is unclear which processes operate for which species and what the efficiencies are, the observed gas composition cannot directly be linked to the ice composition of planet forming bodies. 

The results presented here show that complex molecules can thermally desorb in disks around young stars that have recently undergone an accretion burst. Moreover, their emission extends out to more than 100 AU around V883 Ori and is readily detected and spatially resolved with only one minute of integration with ALMA. V883 Ori is the longest-lasting known outburst and one of, if not the, most luminous. This makes it an archetype for understanding disk chemistry in fainter outbursts. It also provides a look at how younger outbursts may evolve over the century following their outbursts. Young disks like V883 Ori thus provide the unique opportunity to study the chemical complexity at the onset of planet formation.

\acknowledgments

We thank the referee for helpful comments. Astrochemistry in Leiden is supported by NOVA, KNAW and EU A-ERC grant 291141 CHEMPLAN. M.L.R.H. acknowledges support from a Huygens fellowship from Leiden University. J.J.T. acknowledges support from the Homer L. Dodge Endowed Chair at the University of Oklahoma and NWO grant 639.041.439. This paper makes use of the following ALMA data: ADS/JAO.ALMA\#2013.1.00710.S, and \\ ADS/JAO.ALMA\#2015.1.00350.S. ALMA is a partnership of ESO (representing its member states), NSF (USA) and NINS (Japan), together with NRC (Canada), MOST and ASIAA (Taiwan), and KASI (Republic of Korea), in cooperation with the Republic of Chile. The Joint ALMA Observatory is operated by ESO, AUI/NRAO and NAOJ.

\end{document}